\documentclass[12pt,british,english,pof,aip,reprint,amsmath,amssymb]{revtex4-1}
\usepackage[T1]{fontenc}
\usepackage[latin9]{inputenc}
\usepackage{textcomp}
\usepackage{amsmath}
\usepackage{amssymb}
\usepackage{graphicx}
\usepackage{esint}

\makeatletter
 
 \@ifundefined{textcolor}{}
 {%
   \definecolor{BLACK}{gray}{0}
   \definecolor{WHITE}{gray}{1}
   \definecolor{RED}{rgb}{1,0,0}
   \definecolor{GREEN}{rgb}{0,1,0}
   \definecolor{BLUE}{rgb}{0,0,1}
   \definecolor{CYAN}{cmyk}{1,0,0,0}
   \definecolor{MAGENTA}{cmyk}{0,1,0,0}
   \definecolor{YELLOW}{cmyk}{0,0,1,0}
 }

\def\vec#1{\boldsymbol{\mathit{#1}}}
\def\underline#1{\boldsymbol{\mathsf{#1}}}

\makeatother

\usepackage{babel}
\begin{document}

\title{Weakly nonlinear stability analysis of MHD channel flow \\
using an efficient numerical approach}

\author{Jonathan Hagan}

\author{J\={a}nis Priede}

\email{j.priede@coventry.ac.uk}

\selectlanguage{english}%

\affiliation{Applied Mathematics Research Centre, Coventry University, Coventry,
UK}

\date{\today}
\begin{abstract}
We analyze weakly nonlinear stability of a flow of viscous conducting
liquid driven by pressure gradient in the channel between two parallel
walls subject to a transverse magnetic field. Using a non-standard
numerical approach, we compute the linear growth rate correction and
the first Landau coefficient, which in a sufficiently strong magnetic
field vary with the Hartmann number as $\mu_{1}\sim(0.814-\mathrm{i}19.8)\times10^{-3}\textit{Ha}$
and $\mu_{2}\sim(2.73-\mathrm{i}1.50)\times10^{-5}\textit{Ha}^{-4}.$
These coefficients describe a subcritical transverse velocity perturbation
with the equilibrium amplitude $|A|^{2}=\Re[\mu_{1}]/\Re[\mu_{2}](\textit{Re}_{c}-\textit{Re})\sim29.8\textit{Ha}^{5}(\textit{Re}_{c}-\textit{Re}),$
which exists at Reynolds numbers below the linear stability threshold
$\textit{Re}_{c}\sim4.83\times10^{4}\textit{Ha}.$ We find that the
flow remains subcritically unstable regardless of the magnetic field
strength. Our method for computing Landau coefficients differs from
the standard one by the application of the solvability condition to
the discretized rather than continuous problem. This allows us to
bypass both the solution of the adjoint problem and the subsequent
evaluation of the integrals defining the inner products, which results
in a significant simplification of the method.
\end{abstract}
\maketitle

\section{Introduction}

A number of fluid flows can become turbulent while being linearly
stable. Some of these flows, such as, for example, plane Couette flow
and circular pipe (Hagen-Poiseuille) flow, are linearly stable at
all velocities, while other may become linearly unstable at higher
velocities. Typical examples of the latter class of flows are plane
Poiseuille flow and its magnetohydrodynamic counterpart, Hartmann
flow, which arises when a conducting liquid flows in the presence
of a transverse magnetic field. Theoretically, the former is known
to be linearly stable up to the critical Reynolds number $\textit{Re}_{c}=5722.22,$
\citep{Orszag-71} however, experimentally it has been observed to
become turbulent at Reynolds numbers as low as $10^{3}.$ \citep{Carlson-etal-82,Nishioka-Asai-85,Alavyoon-etal-86}
Similarly, Hartmann flow becomes linearly unstable at the local critical
Reynolds number based on the Hartmann layer thickness $R{}_{c}\approx50\,000,$\citep{Lock55,Takashima-96}
whereas turbulence in this flow can be observed at Reynolds numbers
as low as $R_{t}\approx400.$\citep{MorAlb04,Krasnov-etal04,Krasnov-etal13}

Such a subcritical instability can be accounted for by positive feedback
of the perturbation amplitude on its growth rate, which is a non-linear
effect. Thus, a perturbation with sufficiently large amplitude can
acquire positive growth rate at subcritical Reynolds numbers, where
all small-amplitude perturbations are linearly stable. For small-amplitude
perturbations in the vicinity of linear stability threshold this effect
is described by the so-called Landau (Stuart-Landau) equation\citep{Landau-44,Landau-87}.
Whether an instability is sub- or supercritical is determined by the
coefficients of this equation, which are referred to as Landau coefficients
and have to be determined for each particular case. 

It was first suggested by \citet{Lock55} that the instability of
Hartmann flow may be due to finite-amplitude disturbances. This conjecture
was supported by weakly nonlinear stability analysis of a physically
similar asymptotic suction boundary layer.\citep{Hocking-75,Likhachev-76}
Later the same was found to be the case also for the Hartmann boundary
layer.\citep{MorAlb03}

Alternative explanations for the transition to turbulence in Hartmann
flow are based on the energy stability and transient growth theories.
Although the former applies to arbitrary disturbance amplitudes, it
is essentially an amplitude-independent and, thus, linear approach.
Namely, the nonlinear term drops out of the disturbance energy balance
because it neither produces nor dissipates the energy. Using this
approach \citet{Ling-Albo-99} found the Hartmann layer to be energetically
stable when $R_{e}\lesssim26$, which ensures monotonic decay of all
disturbances. This threshold is almost by an order of magnitude lower
than that observed experimentally. As demonstrated in the numerical
study by \citealp{Krasnov-etal04}the optimal transient growth mechanism,
which has been studied for both the Hartmann boundary layer\citep{GerVar-02}
and for the whole Hartmann flow\citep{AirCas04}, is also linear.
As pointed out by \citet{Wal95}, transition to turbulence is mediated
by nonlinear unstable equilibrium states which are not directly related
to the non-normality of the linearized problem responsible for the
transient growth.

The basic formalism of weakly nonlinear stability analysis of plane
Poiseuille flow by the method of amplitude expansion was introduced
by \citet{Stuart-60} and \citet{Watson-60} and later modified by
\citet{Reynolds-Potter-67}. Higher-order Landau coefficients for
plane Poiseuille flow driven by a fixed pressure gradient were computed
by \citet{Sen-Venkateswarlu-83} using both aforementioned methods,
which were found to perform comparably well at supercritical $\textit{Re}$
but not in the subcritical range, where Watson's method encounters
singularities. Asymptotic expansion methods for weakly nonlinear stability
analysis have been reconsidered and surveyed by \citet{Herbert-83},
and substantially extended by \citet{Stewartson-Stuart-71} who used
the method of multiple scales to include slow spatial variation, which
resulted in the complex Ginzburg-Landau equation.\citep{Aranson-Kramer-02}
The method of multiple scales was shown to be equivalent to that of
amplitude expansion \citep{Fujimura-89} as well as to that the center
manifold reduction, which is another technique for deriving the Landau
equation.\citep{Fujimura-91}

The evaluation of Landau coefficients required in weakly nonlinear
stability analysis is technically complicated by the necessity to
solve the adjoint problem and the subsequent evaluation of complex
inner product integrals containing the adjoint eigenfunction. In this
paper, we employ a non-standard approach which is significantly simpler
than the commonly used one.\citep{Huerre-Rossi-98,Schmid-Henningson-01,Yaglom-12}
Our method is based on the application of the solvability condition
to the discretized rather than continuous problem. This allows us
to evaluate Landau coefficients without using the adjoint eigenfunction,
which in our approach is replaced by the left eigenvector. Such a
possibility has been briefly discussed by \citet{Crouch-Herbert-93}
and a similar approach based on Gaussian elimination has been noticed
also by \citet{Sen-Venkateswarlu-83}. 

The paper is organized as follows. In the section to follow, we formulate
the problem and consider general 2D traveling-wave solution, which
is then expanded in small perturbation amplitude to obtain usual expressions
for Landau coefficients. Section \ref{sec:numerics} presents a detailed
development of our approach for Chebyshev collocation method. The
method is validated in Sec. \ref{sec:valid} by computing Landau coefficients
for plane Poiseuille flow driven either by fixed pressure gradient
or flow rate. Section \ref{sec:Results} presents numerical results
concerning both linear and weakly nonlinear stability of Hartmann
flow. The paper is concluded by a summary of results in Sec. \ref{sec:conclusion}.

\section{\label{sec:problem}Formulation of problem }

\begin{figure}
\begin{centering}
\includegraphics[bb=80bp 100bp 380bp 280bp,clip,width=0.5\textwidth]{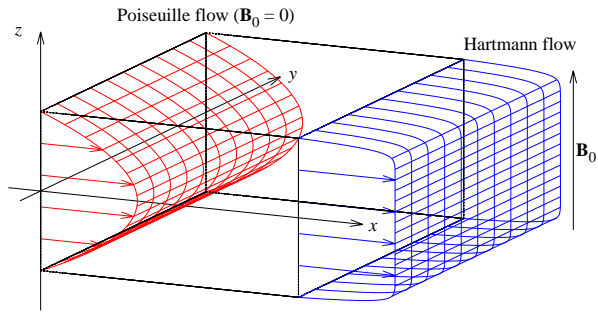} 
\par\end{centering}

\caption{\label{fig:sketch}Sketch of the problem showing velocity profiles
of Poiseuille and Hartmann flows.}
\end{figure}

Consider a flow of incompressible viscous electrically conducting
liquid with the density $\rho,$ kinematic viscosity $\nu$ and electrical
conductivity $\sigma$ driven by a constant gradient of pressure $p$
in the channel of the width $2h$ between two parallel walls in the
presence of a transverse homogeneous magnetic field $\vec{B}$. The
velocity distribution of the flow, $\vec{v}$, is governed by the
Navier-Stokes equation
\begin{equation}
\partial_{t}\vec{v}+(\vec{v}\cdot\vec{\nabla})\vec{v}=-\rho^{-1}\vec{\nabla}p+\nu\vec{\nabla}^{2}\vec{v}+\rho^{-1}\vec{f},\label{eq:NS}
\end{equation}
where $\vec{f}=\vec{j}\times\vec{B}$ is the electromagnetic body
force containing the induced electric current $\vec{j},$ which in
turn is governed by Ohm's law for a moving medium 
\begin{equation}
\vec{j}=\sigma(\vec{E}+\vec{v}\times\vec{B}),\label{eq:Ohm}
\end{equation}
where $\vec{E}$ is the electric field in the stationary frame of
reference. The flow is assumed to be sufficiently slow so that the
induced magnetic field is negligible relative to the imposed one.
This supposes a small magnetic Reynolds number $\textit{Re}_{m}=\mu_{0}\sigma v_{0}h\ll1,$
where $\mu_{0}$ is the permeability of vacuum and $v_{0}$ is the
characteristic velocity of the flow. In addition, we assume that the
characteristic time of velocity variation is much longer than the
magnetic diffusion time $\tau_{m}=\mu_{0}\sigma h^{2}.$ This allows
us to use the quasi-stationary approximation leading to $\vec{E}=-\vec{\nabla}\phi,$
where $\phi$ is the electrostatic potential.\citep{Rob67} The velocity
and current satisfy mass and charge conservation $\vec{\nabla}\cdot\vec{v}=\vec{\nabla}\cdot\vec{j}=0.$
Applying the latter to Ohm's law (\ref{eq:Ohm}) yields 
\begin{equation}
\vec{\nabla}^{2}\phi=\vec{B}\cdot\vec{\omega},\label{eq:phi}
\end{equation}
where $\vec{\omega}=\vec{\nabla}\times\vec{v}$ is vorticity. At the
channel walls $S$, the normal $(n)$ and tangential $(\tau)$ velocity
components satisfy impermeability and no-slip boundary conditions
$\left.v_{n}\right\vert _{s}=0$ and $\left.v_{\tau}\right\vert _{s}=0.$
Electrical conductivity of the walls is irrelevant for the type of
flow considered in this study. 

We employ right-handed Cartesian coordinates with the origin set at
the mid-height of the channel, the $x$- and the $z$-axes directed,
respectively, against the applied pressure gradient $\vec{\nabla}p_{0}=P\vec{e}_{x}$
and along the magnetic field $\vec{B}=B\vec{e}_{z}$ so that the channel
walls are located at $z=\pm h,$ as shown in figure \ref{fig:sketch},
and the velocity is defined as $\vec{v}=(u,v,w).$ Subsequently, all
variables are non-dimensionalized by using $h,$ $h^{2}/\nu$ and
$h\nu B$ as the length, time and electric potential scales, respectively.
The velocity is scaled by the viscous diffusion speed $\nu/h,$ which
we employ as the characteristic velocity instead of the commonly used
center-line velocity.

The problem admits a rectilinear base flow 
\begin{equation}
\vec{v}_{0}(z)=\bar{u}_{0}(z)\vec{e}_{x}=\textit{Re}\,\bar{u}(z)\vec{e}_{x}\label{eq:v0}
\end{equation}
 for which Eq. (\ref{eq:NS}) reduces to 
\begin{equation}
\bar{u}''-\textit{Ha}^{2}\bar{u}=\bar{P},\label{eq:bflw}
\end{equation}
where $\textit{Re}=Uh/\nu$ is the Reynolds number based on the center-line
velocity $U,$ $\textit{Ha}=hB\sqrt{\sigma/\rho\nu}$ is the Hartmann
number, and $\bar{P}$ is a dimensionless pressure gradient satisfying
the normalization condition $\bar{u}(0)=1.$ This equation defines
the well-known Hartmann flow profile

\begin{equation}
\bar{u}(z)=\frac{\cosh(\textit{Ha})-\cosh(z\textit{Ha})}{\cosh(\textit{Ha})-1}\label{eq:Haflw}
\end{equation}
with $\bar{P}=-\frac{\textit{Ha}^{2}\cosh(\textit{Ha})}{\cosh(\textit{Ha})-1},$
which relates the center-line velocity with the applied pressure gradient
$P=\bar{P}U\nu\rho/h^{2}.$ In the weak magnetic field $(\textit{Ha}\ll1),$
the Hartman flow reduces to the classic plane Poiseuille flow $\bar{u}(z)=1-z^{2}.$ 

At sufficiently high $\textit{Re},$ the base flow can become unstable
with respect to infinitesimal perturbations $\vec{v}_{1},$ which
due to the invariance of the base flow in both $t$ and $\vec{x}=(x,y)$
can be sought as 
\begin{equation}
\vec{v}_{1}(\vec{r},t)=\vec{\hat{v}}(z)\mathrm{e}^{\lambda t+\mathrm{i}\vec{k}\cdot\vec{x}}+\mbox{c.c.},\label{eq:v1}
\end{equation}
where $\vec{\hat{v}}(z)$ is complex amplitude distribution, $\lambda$
is temporal growth rate, and $\vec{k}=(\alpha,\beta)$ is the wave
vector. The incompressibility constraint, which takes the form $\vec{D}_{1}\cdot\vec{\hat{v}}=0,$
where $\vec{D}_{1}\equiv\vec{e}_{z}\frac{\mathrm{d}\,}{\mathrm{d}z}+\mathrm{i}\vec{k}$
is a spectral counterpart of the nabla operator, is satisfied by expressing
the component of velocity perturbation in the direction of the wave
vector as $\hat{u}_{\shortparallel}=\vec{e}_{\shortparallel}\cdot\vec{\hat{v}}=\mathrm{i}k^{-1}\hat{w}',$
where $\vec{e}_{\shortparallel}=\vec{k}/k$ and $k=|\vec{k}|.$ Taking
the \emph{curl} of the linearized counterpart of Eq. (\ref{eq:NS})
to eliminate the pressure gradient and then projecting it onto $\vec{e}_{z}\times\vec{e}_{\shortparallel},$
after some transformations we obtain the Orr-Sommerfeld equation 
\begin{equation}
\lambda\vec{D}_{1}^{2}\hat{w}=\left[\vec{D}_{1}^{4}-\textit{Ha}^{2}(\vec{e}_{z}\cdot\vec{D}_{1})^{2}+\mathrm{i}k\textit{Re}(\bar{u}''-\bar{u}\vec{D}_{1}^{2})\right]\hat{w},\label{eq:OSm}
\end{equation}
which contains the electromagnetic term proportional to $\textit{Ha}^{2}.$
The no-slip and impermeability boundary conditions require 
\begin{equation}
\hat{w}=\hat{w}'=0\quad\mbox{at}\quad z=\pm1.\label{eq:bc-2}
\end{equation}
The equation above is written in a non-standard form corresponding
to our choice of the characteristic velocity. Note that Reynolds number
appears in this equation as a factor at the convective term rather
than its reciprocal at the viscous term as in the standard form. As
a result, the growth rate $\lambda$ differs by a factor of $\textit{Re}$
from its standard definition. In this form, the equation is slightly
more convenient for the subsequent numerical solution. 

Since the equation above admits Squire's transformation as in the
non-magnetic case,\citep{Drazin-Reid-81} in the following we consider
only two-dimensional perturbations $(k=\alpha),$ which are the most
unstable.\citep{Lock55} The linear stability problem is solved numerically
using a Chebyshev collocation method.\citep{Hagan-Priede-13} Linear
stability analysis yields marginal values of $\textit{Re}$ depending
on $\vec{k}$ for which neutrally stable perturbations defined by
$\Re[\lambda]=0$ are possible. The lowest marginal value of $\textit{Re}$
is the critical Reynolds number $\textit{Re}_{c}.$ For $\textit{Re}>\textit{Re}_{c},$
the linear stability theory predicts exponentially growing perturbations.
Evolution of unstable perturbations depends on the nonlinear effects
which may either inhibit or enhance the growth rate leading, respectively,
to what is known as super- and subcritical instabilities. The former
is expected to set in only at supercritical Reynolds numbers, whereas
the latter can be triggered by sufficiently large amplitude perturbations
also in a certain range of subcritical Reynolds numbers.

\subsection{2D equilibrium states}

In order to determine whether instability is super- or subcritical,
we employ an approach similar to that of \citet{Reynolds-Potter-67},
which is known as the method of ``false problems'',\citep{Joseph-Sattinger-72,Herbert-83}
and search for equilibrium solution in the vicinity of $\textit{Re}_{c}$
as follows. The neutrally stable mode (\ref{eq:v1}) with a purely
real frequency $\omega=-\mathrm{i}\lambda$ interacting with itself
through quadratically nonlinear term in Eq. (\ref{eq:NS}) is expected
to produce a steady streamwise-invariant perturbation of the mean
flow as well as a second harmonic $\sim\mathrm{e}^{2\mathrm{i}(\omega t+\alpha x)}.$
Subsequent nonlinear interactions produce higher harmonics, which
similarly to the fundamental and second harmonics travel with the
same phase speed $c=-\omega/\alpha.$ Thus, the solution can be sought
in the form of traveling waves
\begin{equation}
\vec{v}(\vec{r},t)=\sum_{n=-\infty}^{\infty}E^{n}\vec{\hat{v}}_{n}(z),\label{eq:v2d}
\end{equation}
where $E=\mathrm{e}^{\mathrm{i}(\omega t+\alpha x)}$ contains $\omega,$
which needs to determined together with $\vec{\hat{v}}_{n}$ by solving
a non-linear eigenvalue problem. The reality of solution requires
$\vec{\hat{v}}_{-n}=\vec{\hat{v}}_{n}^{*},$ where the asterisk stands
for the complex conjugate. The incompressibility constraint applied
to the $n$th velocity harmonic results in $\vec{D}_{n}\cdot\vec{\hat{v}}_{n}=0,$
where $\vec{D}_{n}\equiv\vec{e}_{z}\frac{\mathrm{d}\,}{\mathrm{d}z}+\mathrm{i}\vec{e}_{x}\alpha_{n}$
with $\alpha_{n}=\alpha n.$ This constraint can be satisfied by expressing
the streamwise velocity component 
\begin{equation}
\hat{u}_{n}=\vec{e}_{x}\cdot\vec{\hat{v}}_{n}=\mathrm{i}\alpha_{n}^{-1}\hat{w}_{n}'\label{eq:un}
\end{equation}
in terms of the transverse component $\hat{w}_{n}=\vec{e}_{z}\cdot\vec{\hat{v}}_{n},$
which we employ instead of the commonly used stream function. Henceforth,
the prime is used as a shorthand for $\mathrm{d}/\mathrm{d}z.$ Note
that Eq. (\ref{eq:un}) is not applicable to the zeroth harmonic,
for which it yields $\hat{w}_{0}\equiv0.$ Thus, $\hat{u}_{0}$ needs
to be considered separately in this velocity-based formulation. 

Taking the \emph{curl} of Eq. (\ref{eq:NS}) to eliminate the pressure
gradient and then projecting it onto $\vec{e}_{y},$ we obtain 
\begin{equation}
[\vec{D}_{n}^{2}-\mathrm{i}\omega n]\hat{\zeta}_{n}-\textit{Ha}^{2}\hat{u}_{n}'=\hat{h}_{n},\label{eq:vrt}
\end{equation}
where 
\begin{equation}
\hat{\zeta}_{n}=\vec{e}_{y}\cdot\vec{D}_{n}\times\vec{\hat{v}}_{n}=\begin{cases}
\mathrm{i}\alpha_{n}^{-1}\vec{D}_{n}^{2}\hat{w}_{n}, & n\not=0;\\
\hat{u}_{0}', & n=0.
\end{cases}\label{eq:zeta}
\end{equation}
and 
\begin{equation}
\hat{h}_{n}={\displaystyle \sum_{m}}\vec{\hat{v}}_{n-m}\cdot\vec{D}_{m}\hat{\zeta}_{m}\label{eq:h}
\end{equation}
are the $y$-components of the $n$th harmonic of the vorticity $\vec{\zeta}=\vec{\nabla}\times\vec{v}$
and that of the \emph{curl} of the nonlinear term $\vec{h}=\vec{\nabla}\times(\vec{v}\cdot\vec{\nabla})\vec{v}.$
Henceforth, the omitted summation limits are assumed to be infinite.
Separating the terms involving $\hat{u}_{0},$ the sum (\ref{eq:h})
can be rewritten as $\hat{h}_{n}=\mathrm{i}\alpha_{n}^{-1}(\hat{h}_{n}^{w}+\hat{h}_{n}^{u}),$
where 
\begin{eqnarray}
\hat{h}_{n}^{w} & = & n{\displaystyle \sum_{m\not=0}m^{-1}(}\hat{w}_{n-m}\vec{D}_{n}^{2}\hat{w}_{n}'-\hat{w}_{m}'\vec{D}_{n-m}^{2}\hat{w}_{n-m}),\label{eq:hw}\\
\hat{h}_{n}^{u} & = & \mathrm{i}\alpha_{n}[\hat{u}_{0}-\hat{u}_{0}''\vec{D}_{n}^{2}]\hat{w}_{n}\equiv\mathcal{N}_{n}(\hat{u}_{0})\hat{w}_{n}.\label{eq:hu}
\end{eqnarray}
 Eventually, using the expressions above, Eq. (\ref{eq:vrt}) can
be written as 
\begin{equation}
\mathcal{L}_{n}(\mathrm{i}\omega,\hat{u}_{0})\hat{w}_{n}=\hat{h}_{n}^{w},\label{eq:wn}
\end{equation}
with the operator 
\begin{equation}
\mathcal{L}_{n}(\mathrm{i}\omega,\hat{u}_{0})=[\vec{D}_{n}^{2}-\mathrm{i}\omega n]\vec{D}_{n}^{2}-\textit{Ha}^{2}(\vec{e}_{z}\cdot\vec{D}_{n})^{2}-\mathcal{N}_{n}(\hat{u}_{0}).\label{eq:Ln}
\end{equation}
 The equation above governs all harmonics except the zeroth one, for
which it implies $\hat{w}_{0}\equiv0$ in accordance with the incompressibility
constraint (\ref{eq:un}). The zeroth velocity harmonic, which has
only the streamwise component $\hat{u}_{0},$ is governed directly
by the $x$-component of the Navier-Stokes equation (\ref{eq:NS}):
\begin{equation}
\hat{u}_{0}''-\textit{Ha}^{2}\hat{u}_{0}=\hat{P}_{0}+\hat{g}_{0},\label{eq:u0}
\end{equation}
where $\hat{P}_{0}=\bar{P}\textit{Re}$ is a dimensionless mean pressure
gradient and 
\begin{equation}
\hat{g}_{0}=\mathrm{i}\sum_{m\not=0}\alpha_{m}^{-1}\hat{w}_{m}^{*}\hat{w}_{m}''\label{eq:g0}
\end{equation}
is the $x$-component of the zeroth harmonic of the nonlinear term
$\vec{g}=(\vec{v}\cdot\vec{\nabla})\vec{v}.$ Velocity harmonics are
subject to the usual no-slip and impermeability boundary conditions

\begin{equation}
\hat{w}_{n}=\hat{w}_{n}'=\hat{u}_{0}=0\textrm{ at }z=\pm1.\label{eq:bc}
\end{equation}

\subsection{Amplitude expansion}

The equations obtained previously govern equilibrium states of 2D
traveling waves of arbitrary amplitude. In the vicinity of the linear
stability threshold, which represents the main interest here, solution
can be simplified by expanding it in the small perturbation amplitude.
As discussed above, the fundamental mode (\ref{eq:v1}) with amplitude
$O(\epsilon)$ interacting with itself through the quadratically nonlinear
term in Eq. (\ref{eq:NS}) produces a zeroth harmonic, which modifies
the base flow, and a second harmonic. These two harmonics of amplitude
$O(\epsilon^{2})$ further interacting with the fundamental one produce
an $O(\epsilon^{3})$ correction to the latter. The second harmonic
interacting with the fundamental one also gives rise to a third harmonic
with amplitude $O(\epsilon^{3}).$ This perturbation series is represented
by the following expansion:
\begin{equation}
\hat{w}_{n}=\sum_{m=0}^{\infty}\epsilon^{|n|+2m}\tilde{A}^{|n|}|\tilde{A}|^{2m}\hat{w}_{n,|n|+2m},\label{eq:wn-m}
\end{equation}
where $\epsilon\tilde{A}=A$ is an unknown equilibrium amplitude of
the fundamental harmonic and $\tilde{A}=O(1)$ is its normalized counterpart.
The mean flow, which, as mentioned above, needs to be considered separately,
is expanded as
\begin{equation}
\hat{u}_{0}=\hat{u}_{0,0}+\epsilon^{2}|\tilde{A}|^{2}\hat{u}_{0,2}+\ldots.\label{eq:u0-m}
\end{equation}
Similarly, we expand also Reynolds number and the frequency 
\begin{eqnarray}
\textit{Re} & = & \textit{Re}_{0}+\epsilon^{2}\tilde{\textit{Re}}_{2}+\ldots,\label{eq:Re-m}\\
\omega & = & \omega_{0}+\epsilon^{2}\tilde{\omega}_{2}+\ldots,\label{eq:omeg-m}
\end{eqnarray}
where $\textit{Re}_{0}$ is the marginal Reynolds number satisfying
$\Re[\lambda_{0}]=0$ for the mode $\hat{w}_{1,1}$ with the frequency
$\omega_{0}=\Im[\lambda_{0}]$ and the wave number $\alpha;$ $\epsilon^{2}\tilde{\textit{Re}}_{2}=\textit{Re}_{2}$
and $\epsilon^{2}\tilde{\omega}_{2}=\omega_{2}$ are deviations of
the respective quantities from their values at the linear stability
threshold. Substituting these expansions into Eqs. (\ref{eq:wn})
and (\ref{eq:u0}), and collecting terms at equal powers of $\epsilon$
we obtain the following equations. At $O(\epsilon^{0}),$ we have
the base flow equation 
\begin{equation}
\hat{u}_{0,0}''-\textit{Ha}^{2}\hat{u}_{0,0}=-P_{0,0},\label{eq:u00}
\end{equation}
 where $P_{0,0}=\bar{P}\textit{Re}_{0}$ and $\hat{u}_{0,0}=\textit{Re}_{0}\bar{u}(z).$
At $O(\epsilon),$ we recover the Orr-Sommerfeld equation
\begin{equation}
\mathcal{L}_{1}(\mathrm{i}\omega_{0},\hat{u}_{0,0})\hat{w}_{1,1}=0,\label{eq:w11}
\end{equation}
which defines the linear stability threshold. Solution of this eigenvalue
problem for a given wave number $\alpha$ yields $\textit{Re}_{0},$
$\omega_{0}$ and $\hat{w}_{1,1}(z).$ The latter is defined up to
an arbitrary factor which in the non-magnetic case is fixed by the
standard normalization condition 
\begin{equation}
\hat{w}_{1,1}(0)=1.\label{bc:w11}
\end{equation}
At $O(\epsilon^{2}),$ two equations are obtained
\begin{eqnarray}
\hat{u}_{0,2}''-\textit{Ha}^{2}\hat{u}_{0,2} & = & -P_{0,2}-2\alpha^{-1}\Im[\hat{w}_{1,1}^{*}\hat{w}_{1,1}''],\label{eq:u02}\\
\mathcal{L}_{2}(\mathrm{i}\omega_{0},\hat{u}_{0,0})\hat{w}_{2,2} & = & 2[(\hat{w}_{1,1}\hat{w}_{1,1}')'-2\hat{w}_{1,1}'^{2}]',\label{eq:w22}
\end{eqnarray}
which define the mean-flow perturbation $\hat{u}_{0,2}$ and the second
harmonic $\hat{w}_{2,2}$ in terms of $\hat{w}_{1,1}(z).$ The mean-flow
perturbation depends also on the mean pressure gradient perturbation
$P_{0,2},$ which is zero when the flow is driven by a fixed pressure
difference. Alternatively, if the flow rate rather than the pressure
difference is fixed, then $P_{0,2}$ is an additional unknown, which
has to be determined by using the flow rate conservation condition
$\int_{-1}^{1}\hat{u}_{0,2}(z)\, dz=0.$ We start with a fixed mean
pressure gradient corresponding to $P_{0,2}=0.$ In this formulation,
the case of fixed flow rate can readily be reduced to the former by
incorporating $P_{0,2}$ into $\textit{Re}_{2}$ as shown later on. 

To complete the solution we need to proceed to the order $O(\epsilon^{3}),$
which yields 
\begin{equation}
\mathcal{L}_{1}(\mathrm{i}\omega_{0},\hat{u}_{0,0})\hat{w}_{1,3}=\hat{h}_{1,3}^{w}+|A|^{-2}[\mathcal{N}_{1}(\textit{Re}_{2}\bar{u}+|A|^{2}\hat{u}_{0,2})+\mathrm{i}\omega_{2}\vec{D}_{1}^{2}]\hat{w}_{1,1},\label{eq:w13}
\end{equation}
where 
\begin{equation}
\hat{h}_{1,3}^{w}=\frac{1}{2}\left(\hat{w}_{1,1}^{*}\vec{D}_{2}^{2}\hat{w}_{2,2}'-\hat{w}_{2,2}'\vec{D}_{1}^{2}\hat{w}_{1,1}^{*}\right)-\left(\hat{w}_{2,2}\vec{D}_{1}^{2}\hat{w}_{1,1}'^{*}-\hat{w}_{1,1}'^{*}\vec{D}_{2}^{2}\hat{w}_{2,2}\right).\label{eq:hw13}
\end{equation}
Equation (\ref{eq:w13}) defines the correction of the fundamental
harmonic $\hat{w}_{1,3}$ in terms of the lower order perturbations
described above. It is important to notice that the l.h.s. operator
of Eq. (\ref{eq:w13}) is the same as that of the homogeneous Eq.
(\ref{eq:w11}), which is satisfied by $\hat{w}_{1,1}.$ Thus, Eq.
(\ref{eq:w13}) is solvable only when its r.h.s. contains no term
proportional to $\hat{w}_{1,1},$which means that the r.h.s must be
orthogonal to the adjoint eigenfunction $\hat{w}_{1,1}^{+}:$ 
\begin{equation}
\left\langle \hat{w}_{1,1}^{+},\hat{h}_{1,3}^{w}\right.+|A|^{-2}[\mathcal{N}_{1}(\textit{Re}_{2}\bar{u}+\left.|A|^{2}\hat{u}_{0,2})+\mathrm{i}\omega_{2}\vec{D}_{1}^{2}]\hat{w}_{1,1}\right\rangle =0,\label{eq:slvb}
\end{equation}
 where the angle brackets denote the inner product. This solvability
condition leads to the complex frequency perturbation 
\begin{equation}
\mathrm{i}\omega_{2}=\mu_{1}\textit{Re}_{2}+\mu_{2}|A|^{2},\label{eq:omeg2}
\end{equation}
where 
\begin{eqnarray}
\mu_{1} & = & -\left\langle \hat{w}_{1,1}^{+},\mathcal{N}_{1}(\bar{u})\hat{w}_{1,1}\right\rangle ,\label{eq:mu1}\\
\mu_{2} & = & -\left\langle \hat{w}_{1,1}^{+},\mathcal{N}_{1}(\hat{u}_{0,2})\hat{w}_{1,1}+\hat{h}_{1,3}^{w}\right\rangle \label{eq:mu2}
\end{eqnarray}
 for the adjoint eigenfunction normalized as $\left\langle \hat{w}_{1,1}^{+},\vec{D}_{1}^{2}\hat{w}_{1,1}\right\rangle =1.$
Equation (\ref{eq:omeg2}) represents a reduced Landau equation for
the case of equilibrium solution, which requires $\omega_{2}$ to
be real and, thus, yields the sought equilibrium amplitude 
\begin{equation}
|A|^{2}=-\textit{Re}_{2}\Re[\mu_{1}]/\Re[\mu_{2}].\label{eq:a2}
\end{equation}
This amplitude is the same as that resulting from the full Landau
equation with the first Landau coefficient $\mu_{2}$ and the linear
growth rate correction $\mu_{1}\textit{Re}_{2}.$ Note that our non-standard
choice of the characteristic velocity results in the expressions (\ref{eq:mu1})
and (\ref{eq:mu2}) sharing the operator $\mathcal{N}_{1}$ (\ref{eq:hu})
which simplifies numerical evaluation of these expressions. 

The type of instability is determined by the sign of $\Re[\mu_{2}].$
For an instability to be supercritical, which supposes an equilibrium
solution with $|A|^{2}>0$ at positive linear growth rates $\textit{Re}_{2}\Re[\mu_{1}]>0,$
$\Re[\mu_{2}]<0$ is required. Otherwise, instability is subcritical.
In order to calculate the Landau coefficients (\ref{eq:mu1}) and
(\ref{eq:mu2}) following the standard approach outlined above one
needs to solve not only the Orr-Sommerfeld equation (\ref{eq:w11})
but also its adjoint problem for $\hat{w}_{1,1}^{+}$. Both the direct
and adjoint problems, as well as those posed by Eqs. (\ref{eq:u02})
and (\ref{eq:w22}), need to be solved numerically. Then the integrals
in the inner products defining $\mu_{1}$ and $\mu_{2}$ also need
to be evaluated numerically. This standard approach can significantly
be simplified by evading both the solution of the adjoint problem
and the evolution of the inner product integrals. This is achieved
by applying the solvability condition directly to the discretized
problem as demonstrated in the following.

\section{\label{sec:numerics}Numerical method}

The problem will be solved numerically using a Chebyshev collocation
method with the Chebyshev-Lobatto nodes

\begin{equation}
z_{i}=\cos\left(i\pi/N\right),\quad i=0,\cdots,N,\label{zcol}
\end{equation}
at which the discretized solution $(\hat{w}_{n},\hat{u}_{0})(z_{i})=(\underline{w}_{n},\underline{u}_{0})_{i}$
and its derivatives are sought. The latter are expressed in terms
of the former by using the so-called differentiation matrices, which
for the first and second derivatives are denoted by $D_{i,j}^{(1)}$
and $D_{i,j}^{(2)}.$ Explicit expressions of these matrices, which
are too long to presented here, are given by \citet{Peyret-02}. Equations
(\ref{eq:w11}), (\ref{eq:u02}) and (\ref{eq:w22}) are approximated
at the internal collocation points $0<i<N,$ and the boundary conditions
(\ref{eq:bc}) are imposed at the boundary points $i=0,N.$ The operator
$\mathcal{L}_{n}(\mathrm{i}\omega_{0},\hat{u}_{0,0})$ defined by
Eq. (\ref{eq:Ln}), which appears in Eqs. (\ref{eq:w11}) and (\ref{eq:w22})
is represented by the matrix 
\[
\underline{L}_{n}(\mathrm{i}\omega_{0},\underline{u}_{0,0})=\underline{M}_{n}(\underline{u}_{0,0})-\mathrm{i}\omega_{0}\underline{A}_{n},
\]
which contains 
\begin{eqnarray}
\underline{M}_{n}(\underline{u}_{0,0}) & = & \underline{F}_{n}[\underline{A}_{n}^{2}+\textit{Re}_{0}\underline{N}_{n}(\bar{\underline{u}})],\label{eq:Mn}\\
(\underline{A}_{n})_{i,j} & = & (\vec{D}_{n}^{2})_{i,j},\quad0<(i,j)<N,\label{eq:An}
\end{eqnarray}
where the latter represents part of the collocation approximation
of the operator 
\begin{equation}
(\vec{D}_{n}^{2})_{i,j}=D_{i,j}^{(2)}-\alpha_{n}^{2}\delta_{i,j}\label{eq:Dn}
\end{equation}
 related with the internal nodes; $\delta_{i,j}=(\underline{I})_{i,j}$
is the unity matrix. The other matrix in Eq. (\ref{eq:Mn}), 
\begin{eqnarray}
(\underline{N}_{n}(\bar{\underline{u}}))_{i,j} & = & \mathrm{i}\alpha_{n}[\bar{u}_{i}\delta_{i,j}-\bar{u}_{i}''(\underline{A}_{n})_{i,j}],\label{eq:Nm}
\end{eqnarray}
 represents a collocation approximation of the operator (\ref{eq:hu}).
Finally, the factor matrix\citep{Hagan-Priede-13} 
\begin{equation}
\underline{F}_{n}=\underline{I}-\underline{B}_{n}(\underline{C}\underline{A}_{n}^{-1}\underline{B}_{n})^{-1}\underline{C}\underline{A}_{n}^{-1}\label{eq:Fn}
\end{equation}
in Eq. (\ref{eq:Mn}) is due to the no-slip boundary condition $\hat{w}'(\pm1)=0,$
which is represented by $\underline{C}\underline{w}=\underline{0}$
with 
\begin{equation}
C_{ij}=D_{i,j}^{(1)},\quad i=0,N;0<j<N.\label{eq:C}
\end{equation}
It also involves the part of the operator (\ref{eq:Dn}) related with
the boundary nodes: 
\begin{equation}
(\underline{B}_{n})_{i,j}=(\vec{D}_{n}^{2})_{i,j},\quad0<i<N,j=0,N.\label{eq:Bn}
\end{equation}

\begin{figure*}
\centering{}\includegraphics[width=0.5\textwidth]{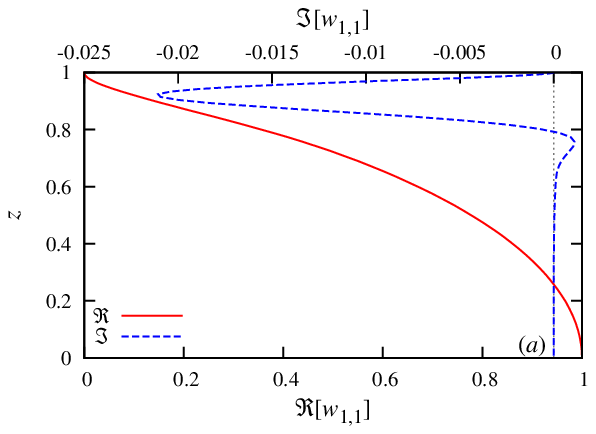}\includegraphics[width=0.5\textwidth]{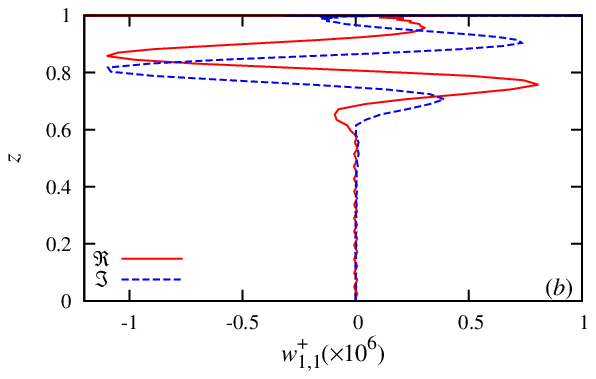}\caption{\label{fig:w11}Real and imaginary parts of the critical perturbation
$\hat{w}_{1,1}$ given by the right eigenvector $\underline{w}_{1,1}$
(a) and those of the respective left eigenvector $\underline{w}_{1,1}^{\dagger}$
(b).}
\end{figure*}

We start with the Orr-Sommerfeld equation, whose collocation approximation
\begin{equation}
\underline{L}_{1}(\lambda,\textit{Re}\bar{\underline{u}})\underline{w}_{1,1}=[\underline{M}_{1}(\textit{Re}\bar{\underline{u}})-\lambda\underline{A}_{1}]\underline{w}_{1,1}=\underline{0},\label{mtr:L1}
\end{equation}
after multiplication by $\underline{A}_{1}^{-1},$ reduces to the
standard complex matrix eigenvalue problem 
\begin{equation}
[\underline{A}_{1}^{-1}\underline{M}_{1}(\textit{Re}\,\bar{\underline{u}})-\lambda\underline{I}]\underline{w}_{1,1}=\underline{0}.\label{mtr:w11}
\end{equation}
The marginal Reynolds number $\textit{Re}_{0}$ for a given wave number
$\alpha$ is determined by the condition $\Re[\lambda_{0}]=0$ for
the eigenvalue $\lambda_{0}$ with the largest real part. Simultaneously
with the right eigenvector $\underline{w}_{1,1},$ we find also the
associated left eigenvector $\underline{w}_{1,1}^{\dagger}.$\citep{Golub-Loan-96}
The right eigenvector is normalized using the condition (\ref{bc:w11}),
and the left one is normalized against the former using the complex
vector dot product $\underline{w}_{1,1}^{\dagger}\cdot\underline{w}_{1,1}=1.$
This normalization simplifies subsequent expressions of Landau coefficients.
Having found $\underline{w}_{1,1}$ we can straightforwardly solve
discretized counterparts of Eqs. (\ref{eq:u02}) and (\ref{eq:w22}),
which yield the mean-flow perturbation $\underline{u}_{0,2}$ and
the complex amplitude distribution of the second harmonic $\underline{w}_{2,2}.$
For the fixed flow rate considered later on, we shall need also the
stream function of the mean-flow perturbation $\underbar{\ensuremath{\psi}}_{0,2},$
which is obtained by solving collocation approximation of $\hat{\psi}_{0,2}'=\hat{u}_{0,2}$
with the symmetry condition $\hat{\psi}_{0,2}(0)=0.$ 

Now, we can proceed to solving our final equation (\ref{eq:w13}),
whose collocation approximation can be written similarly to Eq. (\ref{mtr:L1})
as 
\begin{equation}
\underline{L}_{1}(\mathrm{i}\omega_{0},\textit{Re}_{0}\bar{\underline{u}})\underline{w}_{1,3}=\underline{F}_{1}\underline{h}_{1,3}^{w}+|A|^{-2}[\underline{F}_{1}\underline{N}_{1}(\textit{Re}_{2}\bar{\underline{u}}+|A|^{2}\underline{u}_{0,2})+\mathrm{i}\omega_{2}\underline{A}_{1}]\underline{w}_{1,1},\label{mtr:w13}
\end{equation}
which represents a matrix eigenvalue perturbation problem. For this
system of linear equation to be solvable, its r.h.s multiplied by
$\underline{A}_{1}^{-1},$ as in Eq. (\ref{mtr:w11}), has to be orthogonal
to $\underline{w}_{1,1}^{\dagger}.$\citep{Hinch-71} This discrete
solvability condition leads to the same reduced Landau equation (\ref{eq:omeg2}),
whose coefficients are now defined as
\begin{eqnarray}
\mu_{1} & = & -\underline{w}_{1,1}^{\dagger}\cdot\underline{A}_{1}^{-1}\underline{F}_{1}\underline{N}_{1}(\bar{\underline{u}})\underline{w}_{1,1},\label{eq:mu1-num}\\
\mu_{2} & = & -\underline{w}_{1,1}^{\dagger}\cdot\underline{A}_{1}^{-1}\underline{F}_{1}(\underline{N}_{1}(\underline{u}_{0,2})\underline{w}_{1,1}+\underline{h}_{1,3}^{w}).\label{eq:mu2-num}
\end{eqnarray}
Note that a similar projection onto the left eigenvector of discretized
system is also used to construct reduced models in the flow control
problems\citep{Akervik-etal-07}.

\section{\label{sec:valid}Validation of the method}

In this section, the numerical method will be validated by computing
the first Landau coefficient for plane Poiseuille flow which corresponds
to $\textit{Ha}=0.$ Owing to the symmetry of the problem, both $\hat{w}_{1,1}$
and $\hat{u}_{0,2}$ are even, whereas $\hat{w}_{2,2}$ is an odd
function of $z$. This allows us to search the solution only in the
upper half of the layer which halves the number of required collocation
points. $M=N/2=32$ collocation points in the half-channel is sufficient
to obtain the critical Reynolds number $\textit{Re}_{c}=5772.22$,
frequency $\omega_{c}=-1555.18$ and wave number $\alpha_{c}=1.02055$
to six significant figures.

\begin{figure*}
\centering{}\includegraphics[width=0.5\textwidth]{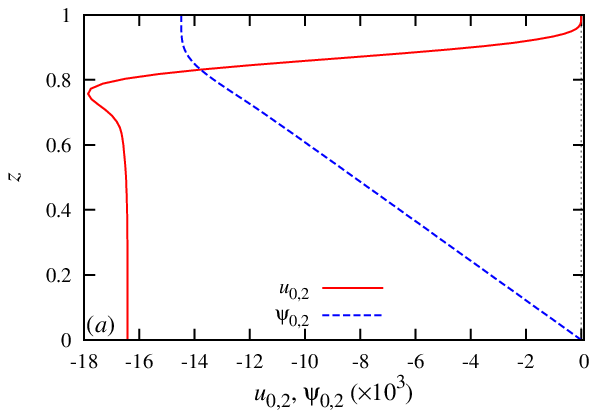}\includegraphics[width=0.5\textwidth]{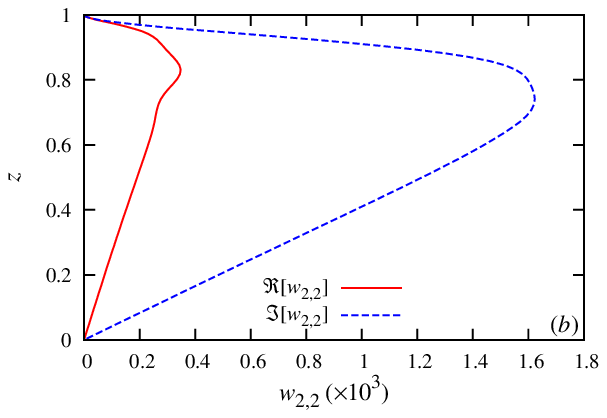}\caption{\label{fig:u02-w22}Velocity $\hat{u}_{0,2}$ and the associated stream
function $\hat{\psi}_{0,2}$ of the mean-flow perturbation (a); the
real and imaginary parts of the second harmonic amplitude $\hat{w}_{2,2}$
(b). }
\end{figure*}

The real and imaginary parts of the critical perturbation $\hat{w}_{1,1},$
which is given by the right eigenvector $\underline{w}_{1,1},$ are
plotted in Fig. \ref{fig:w11} together with the respective left eigenvector
$\underline{w}_{1,1}^{\dagger}.$ Note that the latter is orthogonal
to all other right eigenvectors but $\underline{w}_{1,1},$ and has
only a numerical but no physical meaning. Because of different inner
product definitions for the continuous and discrete problems, $\underline{w}_{1,1}^{\dagger}$
is also distinct from the adjoint eigenfunction $\hat{w}_{1,1}^{+}.$
Distributions of the mean-flow perturbation and that of the complex
amplitude of the second harmonic in the top half of the layer are
plotted in Fig. \ref{fig:u02-w22}. Note that due to the non-standard
scaling, our dimensionless frequency and velocity differ by a factor
of $\textit{Re}_{c}$ from the values obtained with the conventional
scaling based on the center-line velocity.

\begin{figure}
\centering{}\includegraphics[width=0.5\textwidth]{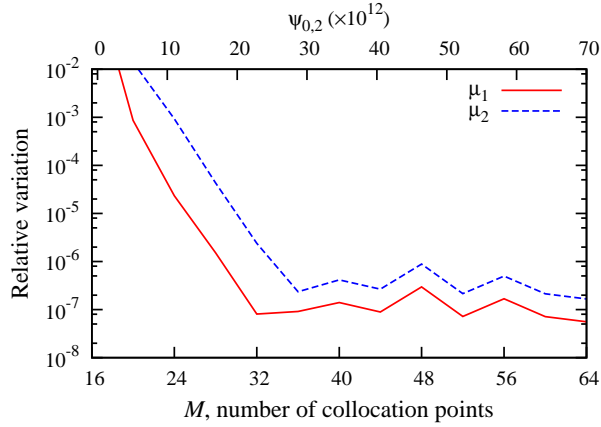}\caption{\label{fig:err}Relative variation of Landau coefficients with the
number of collocation points $M.$ }
\end{figure}

\begin{figure*}
\centering{}\includegraphics[width=0.5\textwidth]{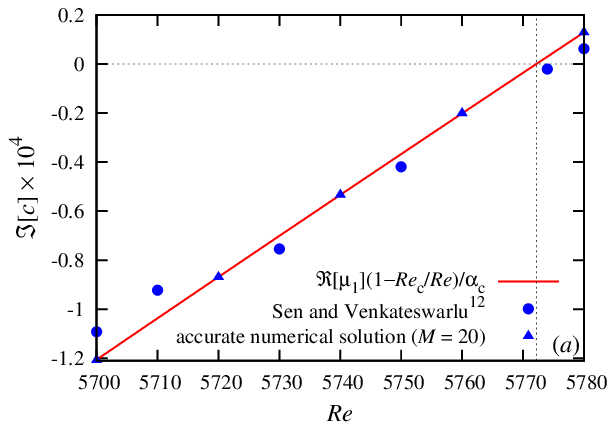}\includegraphics[width=0.5\textwidth]{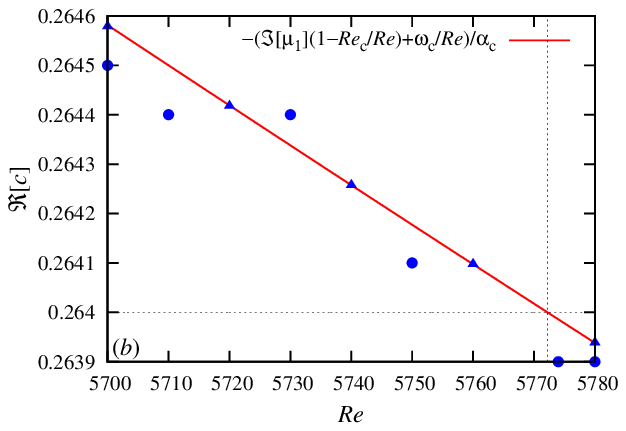}\caption{\label{fig:Landau-c}Imaginary (a) and real (b) parts of the complex
phase velocity $c=-\mathrm{i}\lambda/\textit{Re}\,\alpha$ of the
most unstable mode in the vicinity of the critical Reynolds number
$\textit{Re}_{c}$ calculated using $\mu_{1}$ and supplied by the
linear stability analysis (triangles) and taken from \citet{Sen-Venkateswarlu-83}
(circles). }
\end{figure*}

Substituting the above results into Eqs. (\ref{eq:mu1}) and (\ref{eq:mu2})
we obtain 
\begin{eqnarray*}
\mu_{1} & = & 0.0097118-\mathrm{i}0.222596,\\
\mu_{2} & = & 0.0049382-\mathrm{i}0.0239131.
\end{eqnarray*}
As seen from Fig. \ref{fig:err}, $M\gtrsim32$ collocation points
produce Landau coefficients with about six significant figures. The
first and most important result is $\Re[\mu_{2}]>0,$ which, as discussed
above, confirms the subcritical nature of this instability in agreement
with the previous studies. The linear growth rate coefficient $\mu_{1}$
has been computed explicitly by \citet{Stewartson-Stuart-71}, who
found $d_{1}=(0.17+\mathrm{i}0.8)\times10^{-5}$ for the standard
normalization. Rescaling our result with the center-line velocity,
we obtain $\tilde{\mu}_{1}=\mu_{1}/\textit{Re}_{c}=(0.168251-\mathrm{i}3.85633)\times10^{-5},$
whose real part is close to that of $d_{1},$ while the imaginary
part is significantly different. The reason for this difference is
unclear. In addition, $\mu_{1}$ can be verified against the numerical
results of linear stability analysis for the complex growth rate in
the vicinity of the linear stability threshold, where $\delta\lambda=\lambda-\lambda_{c}\approx\mu_{1}(\textit{Re}-\textit{Re}_{c}).$
As seen in Fig. \ref{fig:Landau-c}, the complex phase speed $c=-\mathrm{i}\lambda/\textit{Re}\,\alpha,$
which is commonly used instead of $\lambda,$ is accurately reproduced
by $\mu_{1}$ in the vicinity of $\textit{Re}_{c}.$ This confirms
the accuracy of $\mu_{1}$ found above.

In order to compare our Landau coefficient $\mu_{2}$ with previous
results, we have to take into account not only our non-standard normalization
but also that $A$ in our case stands for the amplitude of the transverse
velocity component $w,$ whereas in previous studies it denotes the
amplitude of the stream function $\psi,$ which is related to the
former by $\hat{w}=-\mathrm{i}\alpha\hat{\psi}.$ Thus, our $\mu_{2}$
rescales as 
\[
\tilde{\mu}_{2}=\mu_{2}\alpha_{c}^{2}\textit{Re}_{c}=29.659-\mathrm{i}143.622.
\]
This result is close to $\tilde{\mu}_{2}=\mathrm{i}\alpha_{c}K_{1}=29.46-\mathrm{i}143.41$
found by \citet{Sen-Venkateswarlu-83} using the method of \citet{Reynolds-Potter-67}
for $\textit{Re}_{c}=5774,$ $\alpha_{c}=1.02$ and $c_{r}=0.2639.$
Note that $K_{1}$ is mistaken for $\tilde{\mu}_{2}$ by \citet{Schmid-Henningson-01},
who denote it by $\lambda_{2}.$ 

\citet{Reynolds-Potter-67} used their original method of ``false
solution'' to obtain the first relatively accurate values of Landau
coefficients for fixed flow rate. Our solution obtained for fixed
pressure gradient can easily be converted into that for fixed flow
rate by using the non-zero pressure gradient correction $P_{0,2}$
in Eq. (\ref{eq:u02}). As seen from Eq. (\ref{eq:u00}), this correction,
which affects only the magnitude of the base flow, is equivalent to
substituting $\textit{Re}_{2}$ by 
\[
\textit{Re}_{2}^{q}=\textit{Re}_{2}+|A|^{2}P_{0,2}/2.
\]
Requiring the pressure correction $P_{0,2},$ which according to the
expression above produces a flow rate perturbation $|A|^{2}P_{0,2}\bar{\psi}(1),$
to compensate $2|A|^{2}\hat{\psi}_{0,2}(1),$ which is the flow rate
perturbation at fixed pressure gradient, we obtain 
\[
P_{0,2}/2=-\hat{\psi}_{0,2}(1)/\bar{\psi}(1)=0.00217238,
\]
where $\bar{\psi}(1)=\int_{0}^{1}\bar{u}(z)\, dz=\frac{2}{3}.$ Thus,
the substitution of $\textit{Re}_{2}$ by $\textit{Re}_{2}^{q}$ in
Eq. (\ref{eq:omeg2}) results in the replacement of $\mu_{2}$ by
\[
\mu_{2}^{q}=\mu_{2}+\mu_{1}P_{0,2}/2=0.0051492-\mathrm{i}0.0287487.
\]
Rescaling $\mu_{2}^{q}$ with the critical Reynolds number based on
the mean velocity $\bar{\textit{Re}}_{c}=\frac{2}{3}\textit{Re}_{c}=3848.08$
and the critical wave number $\alpha_{c}=1.02071,$ which are the
values used by \citet{Reynolds-Potter-67}, we have
\[
\bar{\mu}_{2}^{q}=\mu_{2}^{q}\alpha_{c}^{2}\bar{\textit{Re}}_{c}=20.64-\mathrm{i}115.26,
\]
which is close to $\bar{\mu}_{2}^{q}=a^{(2)}+\mathrm{i}b^{(2)}=19.7-\mathrm{i}111$
found by \citet{Reynolds-Potter-67}. 

Alternatively, rescaling $\mu_{2}^{q}$ with $\textit{Re}_{c}$ based
on the center-line velocity and the accurate value of $\alpha_{c},$
we obtain
\[
\tilde{\mu}_{2}^{q}=\mu_{2}^{q}\alpha_{c}^{2}\textit{Re}_{c}=30.957-\mathrm{i}172.83,
\]
which agrees well with $\tilde{\mu}_{2}=30.96126-\mathrm{i}172.8268$
and $\tilde{\mu}_{2}=30.95616-\mathrm{i}172.8335$ obtained respectively
by the amplitude expansion using a highly accurate Chebyshev collocation
method\citep{Fujimura-89} and by the center manifold reduction using
an expansion in linear eigenfunctions.\citep{Fujimura-97}

\section{\label{sec:Results}Results}

\subsection{Linear stability threshold of Hartmann flow}

\begin{figure*}
\begin{centering}
\includegraphics[width=0.5\textwidth]{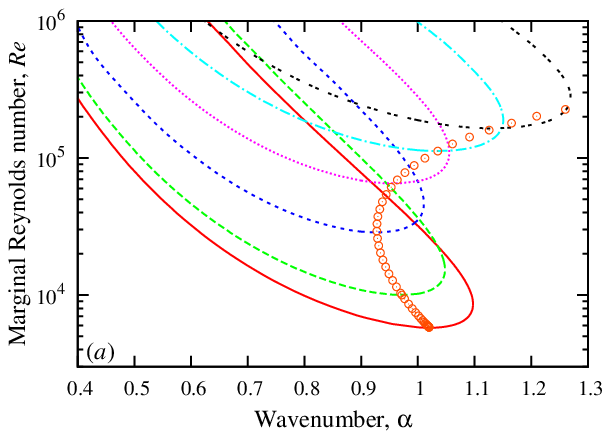}\includegraphics[width=0.5\textwidth]{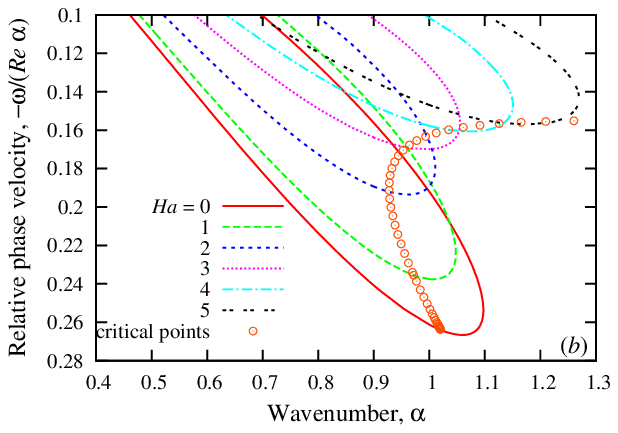}
\par\end{centering}

\caption{\label{fig:rewk}Marginal Reynolds number (a) and the relative phase
velocities of neutrally stable modes (b) against wave number for various
Hartmann numbers. }
\end{figure*}

We start with revisiting the linear stability threshold of the Hartmann
flow which is defined by the marginal Reynolds number at which perturbations
with positive temporal growth rate $\Re[\lambda]$ appear. This Reynolds
numbers and the associated phase velocity of neutrally stable modes
are plotted in Fig. \ref{fig:rewk}(a) versus the wave number $\alpha$
for several Hartmann numbers. The non-magnetic case $(\textit{Ha}=0)$
corresponds to the classic plane Poiseuille flow. First, it is seen
that only the modes with sufficiently small wave numbers can be become
linearly unstable. Second, each such a mode can be linearly unstable
only in a limited range of Reynolds numbers. Namely, besides the lower
marginal Reynolds number by exceeding which mode of a given wave number
turns linearly unstable, there is also an upper marginal Reynolds
number by exceeding which it becomes linearly stable. Linear stability
threshold corresponds the lowest marginal Reynolds number which is
referred to as the critical Reynolds number. For non-magnetic case
$(\textit{Ha}=0)$, the critical Reynolds number is $\textit{Re}_{c}=5772.22,$
and it occurs at the critical wave number $\alpha_{c}=1.02055$.\citep{Orszag-71}
The former is seen in Fig. \ref{fig:rewk}(a) to raise with the Hartmann
number, which means that the flow is stabilized as the magnetic field
is increased. The critical wave number first decreases and then starts
to rise at $\textit{Ha}\gtrsim2.$ 

\begin{figure*}
\begin{centering}
\includegraphics[width=0.5\textwidth]{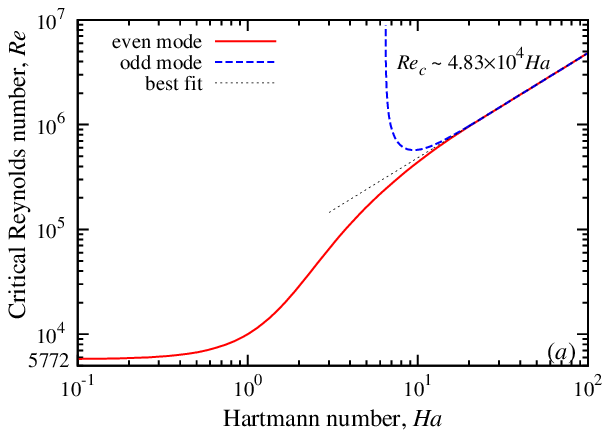}\includegraphics[width=0.5\textwidth]{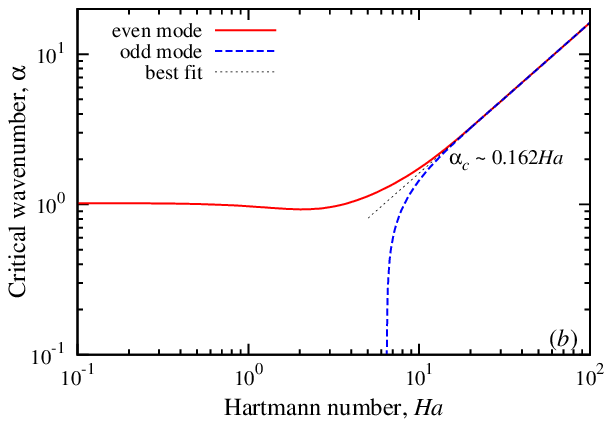}
\par\end{centering}

\begin{centering}
\includegraphics[width=0.5\textwidth]{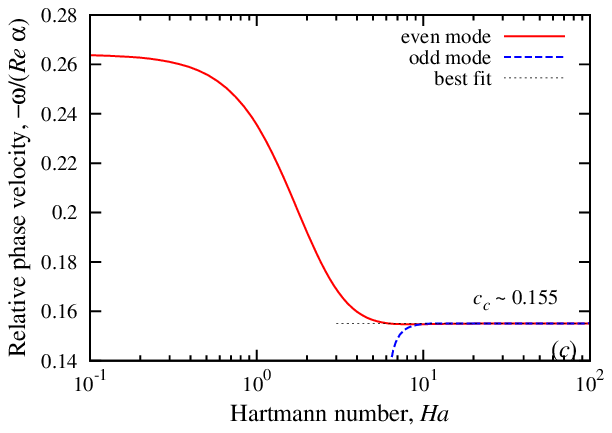}
\par\end{centering}

\caption{\label{fig:crt-Ha} Critical Reynolds number (a), wave number (b)
and phase speed (c) for even and odd instability modes against the
Hartmann number. }
\end{figure*}

\begin{figure*}
\begin{centering}
\includegraphics[bb=0bp 0bp 224bp 154bp,clip,width=0.5\textwidth]{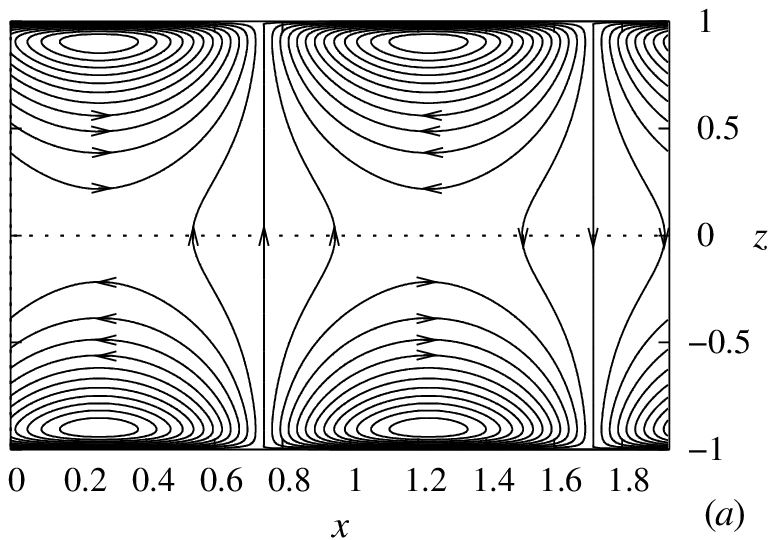}\includegraphics[bb=0bp 0bp 224bp 154bp,clip,width=0.5\textwidth]{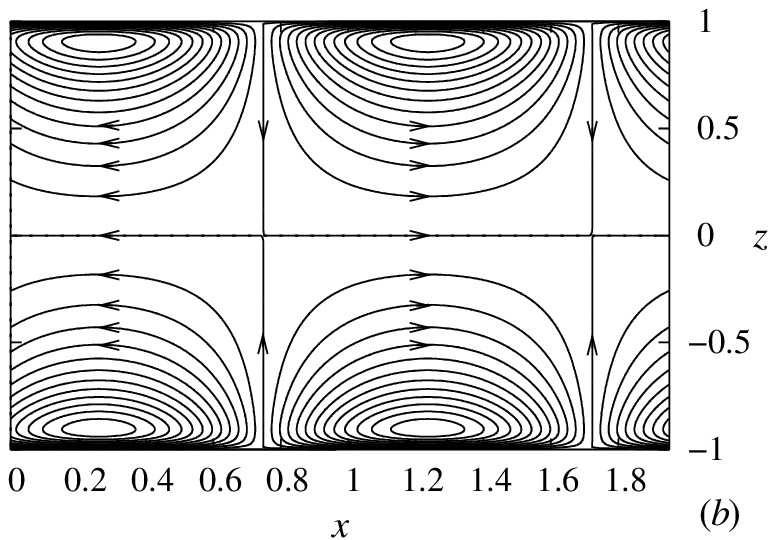}
\par\end{centering}

\caption{\label{fig:slines} Instantaneous streamlines of critical perturbations
for even (a) and odd (b) modes at $\textit{Ha}=20.$ }
\end{figure*}

As seen in Fig. \ref{fig:crt-Ha}, the critical Reynolds number $\textit{Re}_{c}$
and the associated wave number $\alpha_{c}$ both increase in a sufficiently
strong magnetic field $(\textit{Ha}\gtrsim10)$ directly with the
Hartmann number while the relative phase speed $c=-\omega/\textit{Re}\,\alpha$
tends to a constant. The best fit of the numerical results yield 
\begin{eqnarray}
\textit{Re}{}_{c} & \sim & 4.83\times10^{4}\textit{Ha},\label{eq:rec1}\\
\alpha_{c} & \sim & 0.162\textit{Ha},\label{eq:kc1}\\
c_{c} & \sim & 0.155,\label{eq:cc1}
\end{eqnarray}
which agree well with the results of \citet{Takashima-96}. Note that
besides the original instability mode, which develops from the non-magnetic
one, another linearly unstable mode appears at $\textit{Ha}\gtrsim6.5.$
At higher Hartmann numbers, the second mode closely approaches the
original one. Both modes differ by their $z$-symmetry. The transverse
velocity distribution is an even function of $z$ for the former and
an odd function for the latter. This difference becomes unimportant
when $\textit{Ha}\gtrsim20.$ In such a strong magnetic field the
instability becomes localized in the so-called Hartmann boundary layers
of the characteristic thickness 
\begin{equation}
\delta\sim h/\textit{Ha}.\label{eq:Hal}
\end{equation}
First, the localization of instability is implied by the above variations
of $\textit{Re}{}_{c}$ and $\alpha_{c},$ which both become independent
of $\textit{Ha}$ when $\delta$ is used instead of $h$ as the characteristic
length scale. Second, it is also confirmed by the streamline patterns
of the critical perturbations for both modes which are seen in Fig.
\ref{fig:slines} to be very similar to each other. The perturbations
differ by the direction of circulation in the vortices at the opposite
walls, which is the same for the even mode and opposite for the odd
mode. The co-rotating vortices in the even mode are connected through
the mid-plane and, thus, enhance each other, whereas the counter-rotating
vortices in the odd mode tend to suppress each other. In strong magnetic
field, the vortices at the opposite walls become effectively separated
by a stagnant liquid core which makes their interaction insignificant.
This effect has implications for the subsequent weakly nonlinear analysis. 

\subsection{Weakly nonlinear subcritical equilibrium states}

As noted above, the coefficients (\ref{eq:mu1},\ref{eq:mu2}) and,
thus, the equilibrium amplitude (\ref{eq:a2}) determined by them
depend on the normalization of linear eigenfunction. This is because
the equilibrium perturbation (\ref{eq:wn-m}), which is independent
of the normalization, is given by the product of both quantities.
For the classic plane Poiseuille flow, Landau coefficients are usually
calculated by normalizing the linear eigenfunction at the middle of
the layer by the condition (\ref{bc:w11}). This standard normalization,
however, is not suitable for the Hartmann flow. First, it is not compatible
with the odd mode, which satisfies the symmetry condition $\hat{w}_{1,1}(0)=0.$
Second, as discussed above, the same condition is effectively satisfied
also by the even mode when it becomes suppressed in the core of the
layer by a sufficiently strong magnetic field. Thus, instead of the
standard normalization condition (\ref{bc:w11}), we use 
\begin{equation}
\hat{w}_{1,1}''(1)=1,\label{bc:w11-Ha}
\end{equation}
which is related by Eq. (\ref{eq:zeta}) to the vorticity at the wall.
This normalization condition is applicable to both even and odd modes
regardless of the field strength.

\begin{figure*}
\begin{centering}
\includegraphics[width=0.5\textwidth]{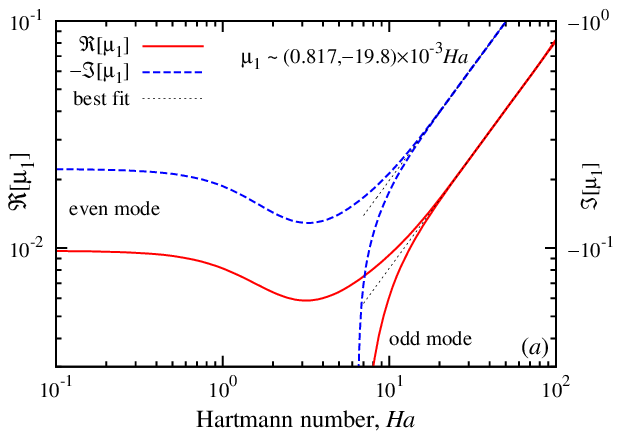}\includegraphics[width=0.5\textwidth]{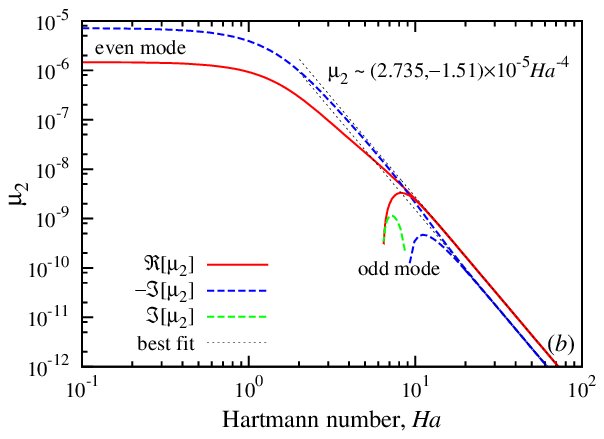}
\par\end{centering}

\caption{\label{fig:Landau} Linear growth rate coefficient $\mu_{1}$ (a)
and and the first Landau coefficient $\mu_{2}$ (b) for odd and even
instability modes normalized with (\ref{bc:w11-Ha}).}
\end{figure*}

The linear growth rate coefficient $\mu_{1}$ and the first Landau
coefficient $\mu_{2}$ computed with this normalization condition
for both critical modes are plotted in Fig. \ref{fig:Landau} against
the Hartmann number. As seen from Eq. (\ref{eq:omeg2}) these coefficients
define the variation of the complex growth rate $\lambda_{2}=\mathrm{i}\omega_{2},$
where $\mu_{1}$ is associated with the deviation of Reynolds number
from its linear stability threshold $\textit{Re}_{2},$ while $\mu_{2}$
accounts for the effect of amplitude $A.$ The real part of $\mu_{1}$
is positive because the critical mode becomes linearly unstable as
$\textit{Re}$ exceeds $\textit{Re}_{c}.$ The positive $\Re[\mu_{2}],$
which is seen in Fig. \ref{fig:Landau}(b) to be the case for all
Hartmann numbers, means that the perturbation amplitude has a positive
feedback on its growth rate. Consequently, the Hartmann flow is sub-critically
unstable regardless of the magnetic field strength. For strong magnetic
field $(\textit{Ha}\gtrsim20)$, the best fit of numerical results
yields 
\begin{eqnarray}
\mu_{1} & \sim & (0.814-\mathrm{i}19.8)\times10^{-3}\textit{Ha},\label{eq:mu1-Ha}\\
\mu_{2} & \sim & (2.73-\mathrm{i}1.50)\times10^{-5}\textit{Ha}^{-4}.\label{eq:mu2-Ha}
\end{eqnarray}
Substituting these asymptotics into Eq. (\ref{eq:a2}) we obtain 
\begin{equation}
|A|^{2}\sim29.8\textit{Ha}^{5}(\textit{Re}_{c}-\textit{Re}).\label{eq:a2-Ha}
\end{equation}
The scaling above is consistent with the relevant length scale of
instability determined by Eq. (\ref{eq:Hal}) which for our choice
of the characteristic velocity $v_{\delta}=\nu/\delta$ leads to $A\sim w''\sim\textit{Ha}^{3}.$
The last result implies that the velocity of equilibrium perturbation
increases asymptotically as $w\sim\textit{Ha},$ which is similar
to the variation of $\textit{Re}_{c}$ with $\textit{Ha}.$ The coefficient
in Eq. (\ref{eq:a2-Ha}) differs from that found by \citet{MorAlb03}
because they normalize the fundamental mode using the velocity maximum,
whereas we use the wall vorticity, which in contrast to the former
is defined explicitly by Eq. (\ref{bc:w11-Ha}).

\begin{figure*}
\begin{centering}
\includegraphics[width=0.5\textwidth]{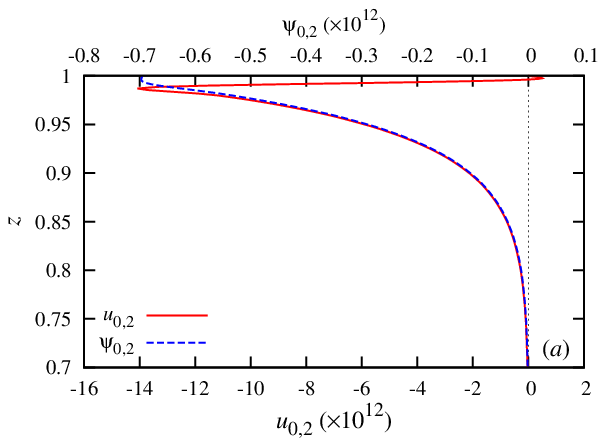}\includegraphics[width=0.5\textwidth]{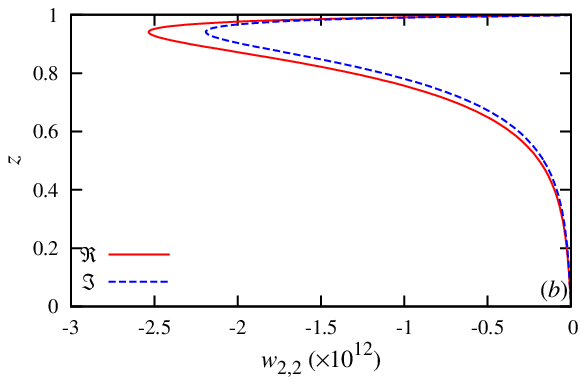}
\par\end{centering}

\centering{}\includegraphics[clip,width=0.5\textwidth]{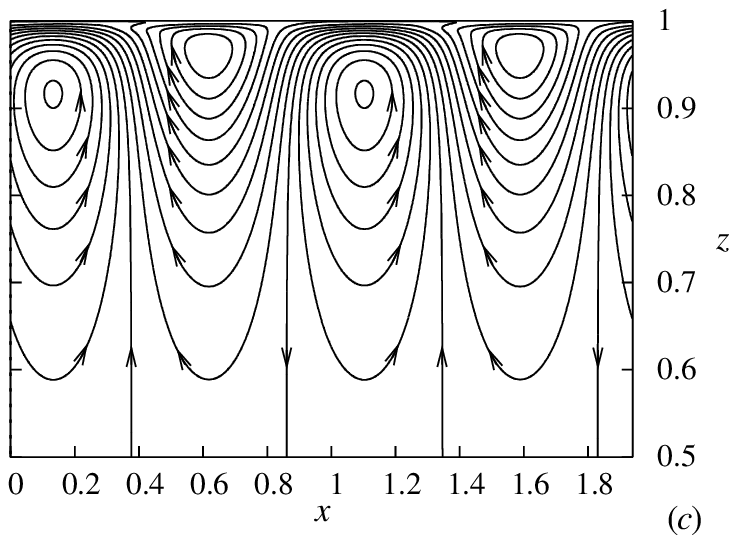}\caption{\label{fig:u02-w22-Ha}Velocity $\hat{u}_{0,2}=\hat{\psi}_{0,2}'$
and the associated stream function $\hat{\psi}_{0,2}$ of the mean
flow perturbation (a); the real and imaginary parts of the second
harmonic amplitude $\hat{w}_{2,2}$ (b), and streamlines of the second-order
perturbation (c) for the even mode at $\textit{Ha}=20.$}
\end{figure*}

The perturbation of the mean flow $\hat{u}_{0,2}(z)$ and the complex
amplitude distribution of the second harmonic $\hat{w}_{2,2}$, which
both are produced by the nonlinear self-interaction of the fundamental
harmonic, are plotted in Figs. \ref{fig:u02-w22-Ha}(a,b). The perturbation
of the flow rate is defined by the stream function $\hat{\psi}_{0,2}(z)=\int_{0}\hat{u}_{0,2}(z)\, dz.$
For strong magnetic field, the best fit yields 
\begin{equation}
\hat{\psi}_{0,2}(1)\sim-4.45\times10^{-5}\textit{Ha}^{-6},\label{eq:psi02}
\end{equation}
whose product with $|A|^{2}$ defined by Eq. (\ref{eq:a2-Ha}) according
to Eq. (\ref{eq:u0-m}) yields the dimensionless perturbation of the
flow rate over half channel. Note that $\textit{Ha}$ cancels out
in this product which is consistent with the dimensional arguments
considered in the paragraph above. Similarly, one can define stream
functions for higher harmonics which satisfy $w_{n}=-\partial_{x}\psi_{n}$
and, thus, lead to the following simple expressions for the complex
amplitudes $\hat{\psi}_{n}=\mathrm{i}\alpha_{n}^{-1}\hat{w}_{n}.$
The streamlines of the second-order perturbation given by $\hat{\psi}_{0,2}(z)-\alpha_{c}^{-1}\Im[\hat{w}_{2,2}(z)\mathrm{e}^{\mathrm{i}2\alpha_{c}x}]$
are shown in Fig. \ref{fig:u02-w22-Ha}(c) for the even mode near
the upper wall at $\textit{Ha}=20.$ Note that the mean-flow perturbation
at fixed pressure gradient reduces the total flow rate by the amount
defined by Eq. (\ref{eq:psi02}). This reduction appears in Fig. \ref{fig:u02-w22-Ha}(c)
as the band of open streamlines undulating between the opposite vortices.
The resulting equilibrium perturbation is formed by the superposition
of this second-order perturbation with the amplitude $|A|{}^{2},$
which is defined by Eq. (\ref{eq:a2-Ha}), and the critical perturbation
with the amplitude $A$ and the streamline pattern shown in Fig. \ref{fig:slines}(a).

\section{\label{sec:conclusion}Conclusion}

The present study was concerned with weakly nonlinear stability analysis
of Hartmann flow, which is an MHD counterpart of plane Poiseuille
flow. Using a non-standard but highly accurate and efficient numerical
approach, which was validated on the classical plane Poiseuille flow,
we computed the first Landau coefficient and the linear growth rate
correction which determine weakly nonlinear evolution of finite small-amplitude
disturbances in the vicinity of linear stability threshold. Hartmann
flow was found to remain subcritically unstable in the whole range
of the magnetic field strength. It means that finite amplitude disturbances
can be become unstable at Reynolds numbers below the linear stability
threshold of Hartmann flow. Next step is to determine how far these
2D as well as 3D finite-amplitude equilibrium states, which are expected
to bifurcate from the former extend into the range of subcritical
Reynolds numbers.\foreignlanguage{british}{\citep{Ehrenstein-Koch-91}}
Such states are thought to mediate transition to turbulence in shear
flows and thus may account for the low transition threshold observed
in both experiments and direct numerical simulations.

The method we used for computing Landau coefficients differs from
the standard one by the application of the solvability condition to
the discretized rather than continuous problem. Expanding equilibrium
solution in small perturbation amplitude in the vicinity of the linear
stability threshold, we obtained a matrix eigenvalue perturbation
problem for the transverse velocity component. Solvability of this
problem requires its inhomogeneous term to be orthogonal to the left
eigenvector. This nonstandard approach allowed us to bypass both the
solution of the adjoint problem and the subsequent evaluation of the
integrals defining the inner products, which resulted in a significant
simplification of the method. The simplicity and relative accuracy
of the method makes it potentially extendible to more complicated
problems like that of MHD duct flow whose weakly nonlinear stability
characteristics are still unclear.\citep{Pot07,PAM10}
\begin{acknowledgments}
J.H. thanks the Mathematics and Control Engineering Department at
Coventry University for funding his studentship.\end{acknowledgments}

\end{document}